 \documentstyle[11pt]{article} 
\begin{document}
\begin{titlepage}
\title{Permanent State Reduction: Motivations, Results, and By-Products
\thanks{\it Talk given at the Int. Symp. on Quantum Communication and
Measurement, Nottingham, 11-17 July 1994.}}
\author{Lajos Di\'osi
\thanks{E-mail: diosi@rmki.kfki.hu}\\
KFKI Research Institute for Particle and Nuclear Physics\\
H-1525 Budapest 114, POB 49, Hungary\\\\
{\it bulletin board ref.: gr-qc/9408007}}
\maketitle
\begin{abstract}
We enlist various motivations underlying the idea that
von Neumann collapses occure
permanently, spontaneously and even universally.
In addition to the conceptual value of permanent reduction theory,
we mention its important by-product promoting new quantum Monte-Carlo
simulations of irrevesible quantum processes, especially in laser optics.
\end{abstract}
\end{titlepage}

\section{Introduction and Contents}

In the last two decades,
serious efforts were concentrated
to find the proper mathematical formalism unifying the
unitary evolution with permanent collapses. Up to now, things
have become well understood at least in the Markovian regime
though the universality of permanent reduction has mere\-ly been modeled on
heuristic grounds.
We attempt to classify main contributions that have led to our present-day
theory of permanent state reduction.
We also emphasize that,
as for the fundamental purpose of the project of permament
reduction, we see a spectacular convergence between it, on one hand,
and the so-called consistent history concept, on the other hand. This
is not very surprising since both projects were initiated to eliminate
the fundamental dichotomy of the ordinary quantum theory.
\bigskip

\leftline{Contents:}
\leftline{1 C-numbers from $\psi$: Reduction}
\leftline{2 Permanent Reduction---Eq. for $\psi(t)$}
\leftline{3 Permanent Reduction---Eqs. for $\psi(t)$ and $z(t)$}
\leftline{4 Consistent Histories}
\leftline{5 By--Products and Conclusion}

\bigskip
\section{C-numbers from $\psi$: Reduction}

However well is quantum mechanics understood its
famous dichotomy has remained unresolved after all.
The quantum state $\psi$ evolves
according to the Schr\"odinger equation:
\begin{equation}
\dot\psi=-iH\psi.
\end{equation}
The knowledge of $\psi$ in itself does not still provide suitable
experimental "c-number" predictions.
One has to turn to the von Neumann reduction theory.
An apparatus interacts with the system in question and changes the
original state of it:
\begin{equation}
\psi\rightarrow\psi_z
\end{equation}
with probability $p_z=|\langle\psi_z|\psi\rangle|^2$,
where $z$ is the c-number recorded by the apparatus.
The possible final states $\{\psi_z\}$ form a complete orthogonal system.
The reduction process above is nonlinear whereas the initial and final
statistical (density) operators are related linearily:
\begin{equation}
\rho\rightarrow\sum_z P_z\rho P_z
\end{equation}
with projectors $P_z=|\psi_z\rangle\langle\psi_z|$.

Let us see what we have done starting with the general ambition to
eliminate the above fundamental dichotomy of the standard quantum mechanics.
What have we discovered for our original purpose and what for others ?

\bigskip
\section{Permanent Reduction---Eq. for $\psi(t)$}

Jumps (2) are not acceptable in Nature, let us interpolate them
by equivalent processes. This was the main motivation to construct
the detailed time evolution equation for $\psi(t)$ interpolating
the von Neumann reduction (2). Let us see the basic results in historical
order.

Bohm and Bub \cite{BohBub66}
proposed deterministic evolution equation where randomness
were only put into initial conditions. Nonlinear Wiener
process was introduced by Pearle, later he found a
a clever gambling game analogy, too \cite{Pea82}.
Gisin \cite{Gis84} pointed out that the stochastic
average of $|\psi(t)\rangle\langle\psi(t)|$, i.e. $\rho(t)$,
must all time obey to linear master equation, and he constructed
the first such nonlinear Wiener process.
Di\'osi \cite{Dio88} showed how a unique nonlinear Wiener
process follows from an arbitrarily given master equation of Lindblad
class. And finally, Gisin and Percival \cite{GisPer9294}
presented the Ito equations of the unique
Wiener process. This has been known as Quantum State Diffusion (QSD) theory
(see Appendix A).

In addition the the first mentioned motivation, it is worthwhile to invoke
two further ones.

{\it Environmental Reduction.}
In everyday experiences, Nature shows spontaneous classicality.
Logically, all c-numbers emerge in corresponding von Neumann reductions.
Macroobjects are never isolated and their
environment acts formally as apparatus \cite{Zeh70}.
This action is permanent and spontaneous.
The macrosystem's density operator $\rho(t)$ is uniquely calculable,
at least in principle.
The stochastic process for the
unraveling $\rho(t)$ into pure state stochastic process $\psi(t)$
can be done in many ways. The QSD theory yields a unique one.

{\it Universal Reduction.}
On philosophical grounds we can hardly accept that classicality is only
due to the particular, e.g., thermal environment. A certain
--- dynamically not specified --- universal "environment" is to be assumed
that guarantees the observed generality and universality
of permanent reduction. Then, postulating the form of the ensemble
evolution equation, the pure state evolution, representing classicality,
follows from QSD theory automatically. For the state of art, see Appendix B.

\bigskip
\section{Permanent Reduction---Eqs. for $\psi(t)$ and $z(t)$}

A typical class of quantum measuring apparatuses
act permanently, i.e., perform unsharp continuous detection.
(Also universal permanent reduction can be represented formally as if
Nature would apply certain apparatuses performing permanent detection.)
The joined statistics of the continuously reduced state $\psi(t)$ and
of the classical record $z(t)$ represents both practical and fundamental
interests. The task that follows is twofold. Firstly, one has to construct
plausible mathematical model of continuous (unsharp) von Neumann reductions
and,  secondly, the simplest form of the corresponding stochastic equations
for $\psi(t)$ and $z(t)$ has to be found.

Let us see a deliberate choice of results.
To my knowledge, the idea of
restricting Feynman-path integrals for a tube along $z(t)$ is
due to Mensky \cite{Men79}.
Barchielli {\it et al.} \cite{BarLanPro82,Bar83}
introduced Gaussian "tubes"
and derived generating functionals for the distribution
functions of the processes $\psi(t),z(t)$.
Caves and Milburn \cite{CavMil87} invented the feed-back
mechanism in the path-integral formalism.
Di\'osi \cite{Dio88Ito}
derived 2 separate Ito-equations: one for $\psi(t)$ and one for
$z(t)$. The first one turned out to be the Gisin
nonlinear equation \cite{Gis84}.
Recently, Di\'osi, Gisin, Halliwell, and Percival \cite{Dioetal94} proved
that the standard QSD corresponds
to permanent reduction onto the (approximate) eigenstates of the
Lindblad generators accompanied by a certain feed-back
{\it \`a la} Caves and Milburn \cite{CavMil87}.
The classical record $z(t)$ is given by $z(t)=\langle L\rangle+\dot w(t)$
where $w(t)$ is the same standard (complex) Wiener process that drives
$\psi(t)$ in QSD.

\bigskip
\section{Consistent Histories}

The motivation for the reinterpretation of quantum mechanics in terms of
CH \cite{Gri84,Omn92,GMH93,DowHal92}
is the same that the motivation for the permanent reduction project:
both models explain the spontaneous emergence of the classical from the
quantum. The CH theory constructs families of repeated (or even continous)
reductions, consistent according to classical logics. It is not at all
surprizing that the CH and the permanent reduction
theories are closely related. Actually,
it can be shown that the samples $\psi(t)$ of QSD are histories
\cite{Dioetal94}
and in most cases they are consistent as well. At least in those
cases where the closed quantum mechanical system splits into a reservoir
and a Markovian subsystem in it, the well-developed Ito-calculus of
permanent reduction offers explicite equations for the CH theory, too.

\bigskip
\section{By-Products and Conclusion}

Even if permanent reduction would not have been successful
in clarifying fundamental
problems it could have, nevertheless, initiated a number of more
practical results. Seemingly, the relevance of nonlinear stochastic
wave equations were realized first in the cited works of fundamental
purposes in the eighties.
Nonlinear stochastic equations, based mostly on independent researches
(see, e.g., Refs.~\cite{Bel89,BarBel91,BelSta92,GagWisMil93})
have now become extensively used in quantum optics. It is not
worthless
to note that the full technical apparatus
for wave function Monte Carlo simulations \cite{DalCasMol92}
in any Markovian system
has been granted, e.g., from 1988 \cite{Dio88} or, in terms of QSD,
from the nineties.
Several good examples could be invoked to show
that practical by-products of the researches on permanent reduction might
well be fertilizing other fields such as quantum optics and communication.

\bigskip
\appendix{\bf A. Quantum State Diffusion}

Consider a system whose state is subjected to permanent reduction and
assume that we know the Lindblad master equation governing the
evolution of the density operator:
\begin{equation}
\dot\rho\equiv{\cal L}\rho=
        -i[H,\rho]+L\rho L^\dagger-{1\over2}\{L^\dagger L,\rho\}
\end{equation}
where $L$ denotes the (vector of) Lindblad generator(s). According
to the QSD theory, the following Ito stochastic
equation belongs to the Liouville superoperator ${\cal L}$:
\begin{equation}
\dot\psi=\left({\cal L}|\psi\rangle\langle\psi|\right)\psi
        +\dot w^\ast\left(L-\langle\psi|L|\psi\rangle\right)\psi
\end{equation}
where $w(t)$ is the (vector of) standard complex Wiener process(es):
\begin{equation}
\dot w(t) \dot w^\ast(t')=\delta(t-t')~~,~~~~~\dot w(t) \dot w(t')=0~~.
\end{equation}
The QSD process (5) has, among others, two noticeable properties. From
the viewpoint of fundamental principles: the above QSD is the only
pure state diffusion process free of any additional parameters
or functions once ${\cal L}$ has been defined. It depends only on ${\cal L}$;
it is invariant under the redefinition of the Lindblad generator(s) $L$.
{}From a more practical viewpoint: the above pure state diffusion process
put as much of the dynamics as possible into the nonlinear drift term
hence the stochasticity is kept minimum. In other words, the nonlinear
drift of QSD offers the best nonlinear Schr\"odinger equation to
approximate the short time behaviour of the density operator.
\bigskip
\appendix{\bf B. Universal Permanent Reduction}

A well--known first attempt to model Universal Permanent Reduction is
the GRW-theory \cite{GRW86}.
Let us illustrate the status of Universal Permanent Reduction project
by a model which relates permanent reduction to gravity \cite{Dio89}.
A compact
definition of the model can be achieved if we specify the Lindblad
generators. They are chosen to be proportional to the square root
of the gravitational constant $G$:
\begin{equation}
L_{\bf k}=\sqrt{G\over4\pi k}\sum_m m\exp\left(i{\bf k}{\bf x}_m\right)
\end{equation}
where $m, {\bf x}_m$ are the masses and coordinate operators, resp., of the
objects belonging to our system; summation should be taken for all masses.
The vector $L_{\bf k}$ is labelled by the wavenumber ${\bf k}$.
A short distance cutoff must be applied at $k\approx10^5cm^{-1}$
\cite{GGP90}.
It can be shown that the corresponding QSD equations reduces to the
ordinary quantum mechanics for microscopic masses. For more massive
objects, permanent reduction become effective and plausible
classicality emerges, e.g., in unique positon distributions.
Gravitational permanent reduction will hopefully find new justification
in the frame of quantum cosmology.
\bigskip

This work was supported by the grant OTKA No. 1822/1991.
The author thanks the Symposium organisers for the kind and
generous invitation.
\bigskip

\end{document}